\documentclass[aip,amsmath,amssymb,reprint]{revtex4-1}

\usepackage{graphicx}
\usepackage{dcolumn}
\usepackage{bm}
\usepackage[version=4]{mhchem}
\usepackage[utf8]{inputenc}
\usepackage[T1]{fontenc}
\usepackage{mathptmx}
\usepackage{etoolbox}
\usepackage{xcolor}

\makeatletter
\def\@email#1#2{%
 \endgroup
 \patchcmd{\titleblock@produce}
  {\frontmatter@RRAPformat}
  {\frontmatter@RRAPformat{\produce@RRAP{*#1\href{mailto:#2}{#2}}}\frontmatter@RRAPformat}
  {}{}
}%
\makeatother

\begin{document}

\title{Tuning of Atomic Layer Deposition Pulse Time through Physics-Informed Bayesian Active Learning}

\author{Pouyan Navabi}
\affiliation{Department of Chemistry, University of Illinois Chicago, Chicago, Illinois, 60607, USA}

\author{Christos G. Takoudis}
\email{takoudis@uic.edu}
\affiliation{Department of Chemistry, University of Illinois Chicago, Chicago, Illinois, 60607, USA}
\affiliation{Department of Biomedical Engineering, University of Illinois Chicago, Chicago, Illinois, 60607, USA}
\affiliation{Department of Chemical Engineering, University of Illinois Chicago, Chicago, Illinois, 60607, USA}

\date{\today}

\begin{abstract}
Atomic Layer Deposition (ALD) process development is often hindered by time-consuming and precursor-intensive tuning cycles required to identify saturation conditions. We introduce a physics-informed Bayesian active learning (BAL) framework that autonomously tunes precursor pulse times by integrating a Langmuir adsorption model directly into the Gaussian Process (GP) kernel. A key innovation is a two-stage parameter estimation strategy that decouples noise filtering from physical parameter extraction: the GP first smooths noisy data through standard prediction, then Langmuir parameters are fitted to the noise-filtered GP predictions. This approach effectively separates signal from experimental noise. We evaluate the framework against a standard data-driven GP across four simulated regimes, demonstrating convergence within five iterations, up to fourfold improvement in prediction accuracy, and two to fourfold reduction in precursor usage. Experimental validation using \ce{TiO2} deposition via Tetrakisdimethylamido Titanium (TDMAT) and ozone confirms that the physics-informed model accurately identifies saturation times for high-coverage targets ($\geq$95\%), with observed deviations at lower saturation levels providing valuable insight into non-ideal desorption behaviors.
\end{abstract}

\maketitle

\section{Introduction}

Atomic layer deposition (ALD) is a vapor-phase thin film deposition technique that operates as a specialized form of chemical vapor deposition (CVD).\cite{1} Unlike CVD, where reactants and co-reactants are introduced simultaneously, ALD relies on sequential, self-limiting surface reactions. A typical ALD cycle comprises four consecutive steps: (1) pulsing the precursor into the reaction chamber, (2) purging with an inert carrier gas, (3) introducing the co-reactant, and (4) performing a final purge. This sequence enables precise control of film thickness and conformality at the atomic scale. 

Depositing materials one atomic layer at a time was initially a specialized task for the semiconductor industry,\cite{2, 3, 4} but it has evolved into a fundamental tool for modern manufacturing. Today, ALD is critical across many fields because it solves a unique challenge: coating complex, deep structures with a precision that other methods simply cannot match. From improving the efficiency of solar cells\cite{5, 6, 7} to encapsulating biomedical implants,\cite{8} the self-limiting nature of ALD provides a level of consistency that is essential for high-performance devices. Yet, as these applications expand, there is a growing need to make the process more efficient, specifically by minimizing waste\cite{9} and optimizing how quickly saturation is achieved.

Traditionally, determining this optimal saturation point is a resource-intensive task. It often requires extensive trial-and-error experimental campaigns that consume significant precursor material and machine time. As process complexity increases, these manual optimization methods become a bottleneck. To address this, data-driven approaches are emerging as a powerful alternative.\cite{10} Machine learning algorithms can analyze in situ data to predict saturation behavior with high accuracy, significantly reducing the time and cost required to develop new ALD recipes.\cite{11}

Multiple process parameters critically influence the efficiency and quality of ALD growth.\cite{1} The substrate temperature must be sufficiently high to activate surface reactions and prevent precursor condensation, yet low enough to avoid thermal decomposition. The precursor canister temperature determines vapor pressure and stability, while chamber pressure and carrier gas flow rate affect film uniformity, growth dynamics, and conformality. Another critical set of parameters includes the pulse and purge durations for both precursor and co-reactant, as these directly control material deposition and film growth per cycle. Tuning these parameters is therefore essential to achieve saturated surface reactions efficiently, minimize precursor waste, and ensure reproducible film quality.

Recent studies have highlighted the potential of machine learning to enhance ALD process development.\cite{10} Methods such as neural networks\cite{12, 13, 14} and Bayesian optimization (BO)\cite{15, 16, 17} have been applied to both computational and experimental workflows to extract saturation times, refine pulse sequences, and minimize experimental effort. In computational studies, particularly those coupling machine learning with computational fluid dynamics (CFD), these models have been used to predict surface coverage dynamics and identify optimal process parameters.\cite{18} Experimentally, Bayesian frameworks have been employed to iteratively suggest new process conditions to minimize defects, effectively reducing the number of required experiments.\cite{15, 16} Paulson et al. developed a data-driven Bayesian framework for optimizing pulse and purge times using simulated datasets for \ce{Al2O3}, \ce{TiO2}, and \ce{W},\cite{17} while more recent efforts toward closed-loop optimization have combined in situ characterization with machine learning models. For instance, neural networks have been integrated with in situ quartz crystal microbalance (QCM) measurements and BO to enable real-time optimization of precursor exposure.\cite{20}

However, these approaches typically rely on reaction-specific trained models that must be developed and parameterized for each chemistry. Moreover, neural network-based approaches often function as digital twins or supervisory systems for the experimental setup, requiring simultaneous implementation of both the model and the control system. While effective, such architectures increase complexity and limit generalizability to new precursors or deposition chemistries. Unlike the approach of Paulson et al., which required reaction-specific trained models, a more flexible framework that balances physical insight with data-driven learning could potentially accelerate ALD process development across diverse material systems.

Recent advances in optimization for physical systems have focused on transitioning from black-box to grey-box Bayesian approaches that incorporate domain-specific physical relationships. Physics-informed BO integrates physics-infused kernels into Gaussian process (GP) surrogate models, enabling the framework to leverage both statistical and physical information during exploration.\cite{30} This strategy has proven effective across diverse applications. In materials design, physics-informed kernels enabled data-efficient identification of optimal NiTi shape memory alloy processing parameters to maximize transformation temperature.\cite{31} In computational chemistry, GP regression combined with BO reconstructed free-energy functions from molecular dynamics trajectories, propagating statistical uncertainty to accurately calculate phase diagrams for systems such as lithium across varying temperatures and pressures,\cite{32} and in conformational analysis, a torsion-potential-based kernel with a specialized acquisition function successfully recovered previously missed low-energy molecular conformations in 40\% of tested biologically relevant molecules.\cite{33} These examples demonstrate that embedding physical structure into the learning framework substantially improves sample efficiency and extrapolation capability compared to purely data-driven approaches.

In this study, we present a physics-informed Bayesian active learning (BAL) framework for tuning precursor pulse duration in ALD using an adsorption model based on the Langmuir isotherm. The key methodological innovation is a two-stage parameter estimation strategy. Instead of fitting the Langmuir model directly to sparse, noisy observations, which often yields unstable parameters, we leverage smoothing capabilities of the GP by first generating dense, noise-free predictions from the GP, then fitting the Langmuir parameters to these smoothed predictions. This provides substantially more stable and robust parameter estimates while naturally filtering measurement noise. 

We systematically evaluate our physics-informed active learning framework across four simulated regimes (\textbf{fast saturation}, \textbf{moderate saturation}, \textbf{slow saturation}, and \textbf{high noise}), demonstrating convergence within five iterations, up to fourfold improvement in prediction accuracy, and two to fourfold reduction in precursor usage. Finally, we validate the framework experimentally using in situ spectroscopic ellipsometry feedback for the ALD of \ce{TiO2}. By combining in situ spectroscopic ellipsometry feedback with this physics-informed BAL approach, the system autonomously identifies tuned pulse times with high accuracy, reduces precursor consumption, and accelerates process development.

\section{Methodology}

\subsection{Adsorption Kinetics Models}

\subsubsection{Langmuir Isotherm Model}

We employ a Langmuir isotherm model\cite{21, 22} to represent the adsorption behavior of precursors during ALD with only exposure time as the effective parameter. The rationale for this approach is that in typical ALD processes, the substrate temperature is determined by the chemical thermodynamics of the precursor and substrate, as well as considerations of precursor and substrate stability and the resulting film quality. The canister temperature is set based on the physical properties of the precursor to ensure adequate vapor pressure. Flow rate and chamber pressure are defined by instrument design and growth dynamics. In most ALD cases, these parameters are well-established for a given process, and tuning efforts focus primarily on pulse and purge times.

For chemical adsorption in ALD, the surface reaction can be represented as:

\begin{equation}
\ce{A(g) + S^*(s) <=> AS(s)}
\label{eq:surface_rxn}
\end{equation}
where precursor \ce{A} from the gas phase binds to available surface sites \ce{S^*}, forming the adsorbed species \ce{AS}. Based on this mechanism, we define the kinetics using a modified Langmuir adsorption model. In a real experimental environment, desorption during the purge step is non-negligible.\cite{23} Therefore, the growth rate $y(x)$ as a function of exposure time $x$ is more accurately described as:

\begin{equation}
y(x) = \frac{G_{\max}Kx}{1+Kx} - D(x)
\label{eq:langmuir_desorption}
\end{equation}
where $G_{\max}$ is the maximum surface coverage, $K$ is the equilibrium constant, and $D(x)$ accounts for material desorption. Ideally, $D(x)$ is negligible compared to the adsorption term, and our experimental protocols aim to minimize it.

In our simulations, we assume ideal behavior where $D(x) \approx 0$, effectively setting:

\begin{equation}
y(x) = \frac{G_{\max}Kx}{1+Kx}
\label{eq:langmuir_ideal}
\end{equation}

However, in our experimental validation, we implement a specific protocol to address the desorption term. We employ an extended 600-second purge and measure the thickness after a sequence of four consecutive pulses. The extended purge minimizes the magnitude of the desorption effect for each new parameter set. Furthermore, since typical ALD recipes do not utilize 600-second purges, the prior four-pulse sequence ensures that $D(x)$ effectively becomes a constant offset associated only with the current parameter $x$ being evaluated. This protocol helps isolate the adsorption kinetics necessary for the learning algorithm. In our simulations, the true parameters ($K_{\text{true}}$, $G_{\text{true}}$) are hidden from the physics-informed BAL framework and used only to generate observations.

\subsubsection{Realistic Noise Model}

To simulate realistic experimental limitations, we implement a two-component Gaussian noise model. The first source of uncertainty arises from the thickness measurement method and represents constant absolute measurement noise:

\begin{equation}
\sigma_{y,\text{direct}} = \sigma_{y,\text{abs}}
\label{eq:noise_direct}
\end{equation}
where $\sigma_{y,\text{abs}}$ represents constant absolute detector noise arising from detector resolution, readout noise, and environmental fluctuations.

The second source of noise originates from timing precision limitations of ALD valves. This timing uncertainty propagates through the Langmuir model derivative:

\begin{equation}
\frac{dy}{dx} = \frac{G_{\max}K}{(1+Kx)^2}
\label{eq:langmuir_derivative}
\end{equation}

The contribution of $x$-uncertainty to the observed $y$-noise is obtained from:

\begin{equation}
\sigma_{y,\text{timing}} = \left|\frac{dy}{dx}\right| \times \sigma_{x,\text{abs}} = \frac{G_{\max}K}{(1+Kx)^2} \times \sigma_{x,\text{abs}}
\label{eq:noise_timing}
\end{equation}
where $\sigma_{x,\text{abs}}$ represents constant absolute timing precision. This creates heteroscedastic (input-dependent) noise that is larger at small $x$ where the Langmuir curve has a steeper slope, naturally creating noise that decreases with increasing exposure time.

The total measurement uncertainty combines both independent sources in quadrature:

\begin{equation}
\sigma_{\text{total}}(x) = \sqrt{\sigma_{y,\text{direct}}^2 + \sigma_{y,\text{timing}}^2}
\label{eq:noise_total}
\end{equation}

Observations are then drawn with Gaussian probability as:

\begin{equation}
y_{\text{obs}}(x) = y(x) + \mathcal{N}\left(0, \sigma_{\text{total}}^2(x)\right)
\label{eq:obs_model}
\end{equation}

This noise model is physically motivated by the causality of the experiment where $x$ is the controlled variable and $y$ is the measured variable. Both sources of uncertainty manifest as scatter in the observed $y$-values.

\subsection{Physics-Informed Bayesian Active Learning}

The learning framework serves as a sequential design strategy to efficiently characterize expensive-to-evaluate functions. We employ GP regression to construct a surrogate model of the saturation curve. While standard GP regression is well-documented\cite{24, 25}, our implementation introduces two domain-specific modifications: a physics-informed covariance kernel and a two-stage parameter estimation strategy. All GP regression algorithms and BAL procedures were implemented using Python modules, including scikit-learn\cite{26} for GP regression, scipy\cite{27} for optimization routines, and numpy for numerical computations. The complete implementation code is provided in the supplementary material.

\subsubsection{Gaussian Process Regression}

We model the growth rate $f(x)$ as a Gaussian Process specified by a zero mean function\cite{24} and a covariance kernel $k(x, x')$. Given a set of $n$ observations $D = \{(x_i, y_i)\}_{i=1}^{n}$, the posterior predictive distribution at a new test point $x_*$ is Gaussian $\mathcal{N}(\mu(x_*), \sigma^2(x_*))$, where the mean $\mu(x_*)$ and variance $\sigma^2(x_*)$ are calculated analytically. The 95\% confidence interval (CI) represents the region within $\pm 1.96$ standard deviations of the posterior mean, capturing the uncertainty in the GP predictions. In our active learning framework, we utilize the posterior variance $\sigma^2(x_*)$ to drive exploration via uncertainty sampling, selecting the next sampling point as $x_{\text{next}} = \arg\max_x \sigma^2(x)$.

\subsubsection{Kernel Strategies}

The choice of covariance function (kernel) $k(x, x')$ encodes our prior assumptions about the smoothness and structure of the underlying function. We compare two GP kernel strategies:

\begin{itemize}
\item \textbf{Physics-informed Matérn kernel ($\nu = 2.5$):} integrates a Langmuir transformation into the GP kernel with adaptive physics parameters.
\item \textbf{Pure Matérn kernel ($\nu = 2.5$):} purely statistical GP with no embedded physics.
\end{itemize}

We employ the Matérn kernel rather than other common kernels (such as the squared exponential or radial basis function) because it provides greater flexibility in modeling function smoothness through the parameter $\nu$. The Matérn class encompasses a broad family of covariance functions, and the choice of $\nu = 2.5$ represents a balance between smoothness and flexibility.\cite{28} This value corresponds to functions that are twice differentiable, which aligns well with the expected behavior of adsorption isotherms. Additionally, $\nu = 2.5$ has been found empirically to perform well for physical systems exhibiting smooth but not infinitely differentiable behavior, making it particularly suitable for capturing the characteristic saturation curve of Langmuir kinetics while remaining robust to experimental noise.

\paragraph{Physics-Informed Matérn Kernel}
The physics-informed kernel projects the input $x$ through a Langmuir-inspired transformation:

\begin{equation}
\phi(x) = \frac{G_{\text{phys}}K_{\text{phys}}x}{1+K_{\text{phys}}x}
\label{eq:phi_transform}
\end{equation}

The correlation between $x_i$ and $x_j$ is defined using the Matérn covariance function:

\begin{equation}
k_{\text{phys}}(x_i, x_j) = \frac{2^{1-\nu}}{\Gamma(\nu)} \times \left(\sqrt{2\nu}r\right)^\nu \times K_\nu\left(\sqrt{2\nu}r\right)
\label{eq:matern_phys}
\end{equation}
where $r = |\phi(x_i) - \phi(x_j)|/\ell$, $\ell$ is the correlation length scale (a hyperparameter that controls the distance over which function values are correlated), $\Gamma$ is the gamma function, and $K_\nu$ is the modified Bessel function of the second kind. The length scale $\ell$ is optimized during GP training by maximizing the marginal log-likelihood of the observed data, allowing the model to automatically determine the appropriate correlation structure from the measurements.

A critical innovation in this framework is the estimation of the adaptive parameters $K_{\text{phys}}$ and $G_{\text{phys}}$. Direct fitting of the Langmuir model to sparse, noisy ALD data often yields unstable parameters. Instead, we adopt a hierarchical two-stage approach:

\begin{enumerate}
\item \textbf{Smooth:} The GP is trained on the available noisy data $(x, y_{\text{obs}})$, acting as a sophisticated noise filter that handles heteroscedastic uncertainty.
\item \textbf{Fit:} We generate dense, noise-free predictions from the GP posterior mean on a fine grid and fit the Langmuir parameters $(K_{\text{phys}}, G_{\text{phys}})$ to this smoothed curve.
\end{enumerate}

This decouples noise rejection (handled by the GP) from physical interpretation (handled by the Langmuir fit), resulting in significantly more robust convergence during the early stages of the active learning process.

\paragraph{Pure Matérn Kernel}
A purely data-driven model without physics transformation is defined as:

\begin{equation}
k_{\text{Matérn}}(x_i, x_j) = \frac{2^{1-\nu}}{\Gamma(\nu)} \times \left(\sqrt{2\nu}r'\right)^\nu \times K_\nu\left(\sqrt{2\nu}r'\right)
\label{eq:matern_pure}
\end{equation}
where $r' = |x_i - x_j|/\ell$. The correlation length scale $\ell$ is likewise optimized by maximizing the marginal log-likelihood, providing the model with the flexibility to learn appropriate correlations directly from the data without incorporating domain-specific knowledge. This kernel serves as a baseline that relies solely on data smoothness assumptions without domain-specific structure.

\subsubsection{Active Learning Workflow}

The goal of our methodology is to identify the optimal data points that yield the best Langmuir fit ($K$ and $G_{\max}$ values) and then use this optimized fit to predict the targeted saturation. This methodology ensures that saturation can be accurately predicted even when the targeted saturation falls outside the exploration window. The adjustable parameters for the active learning framework are provided in Table~\ref{tab:params}.

\begin{table}
\caption{\label{tab:params}Adjustable parameters for simulation and Bayesian learning models}
\begin{ruledtabular}
\begin{tabular}{ll}
\textbf{Parameter} & \textbf{Description} \\
\multicolumn{2}{l}{\textit{Simulation Parameters}} \\
$K_{\text{true}}$ & True Langmuir equilibrium constant (1/s) \\
$G_{\max,\text{true}}$ & True maximum surface coverage (\AA) \\
$\sigma_{y,\text{abs}}$ & Absolute $y$-uncertainty (measurement noise, \AA) \\
$\sigma_{x,\text{abs}}$ & Absolute $x$-uncertainty (timing precision, s) \\
$N_{\text{MC}}$ & Number of Monte Carlo runs \\
$N_{\text{iter,MC}}$ & Iterations per Monte Carlo run \\
\multicolumn{2}{l}{\textit{Bayesian Learning Parameters}} \\
$N_{\text{iter}}$ & Number of active learning iterations \\
$N_{\text{survey}}$ & Initial Latin Hypercube Sampling survey points \\
$x_{\min}$ & Exploration window minimum (s) \\
$x_{\max}$ & Exploration window maximum (s) \\
$\nu$ & Matérn smoothness parameter \\
$s_{\text{target}}$ & Target saturation fraction \\
\end{tabular}
\end{ruledtabular}
\end{table}

\begin{figure}
\includegraphics[width=3.375in]{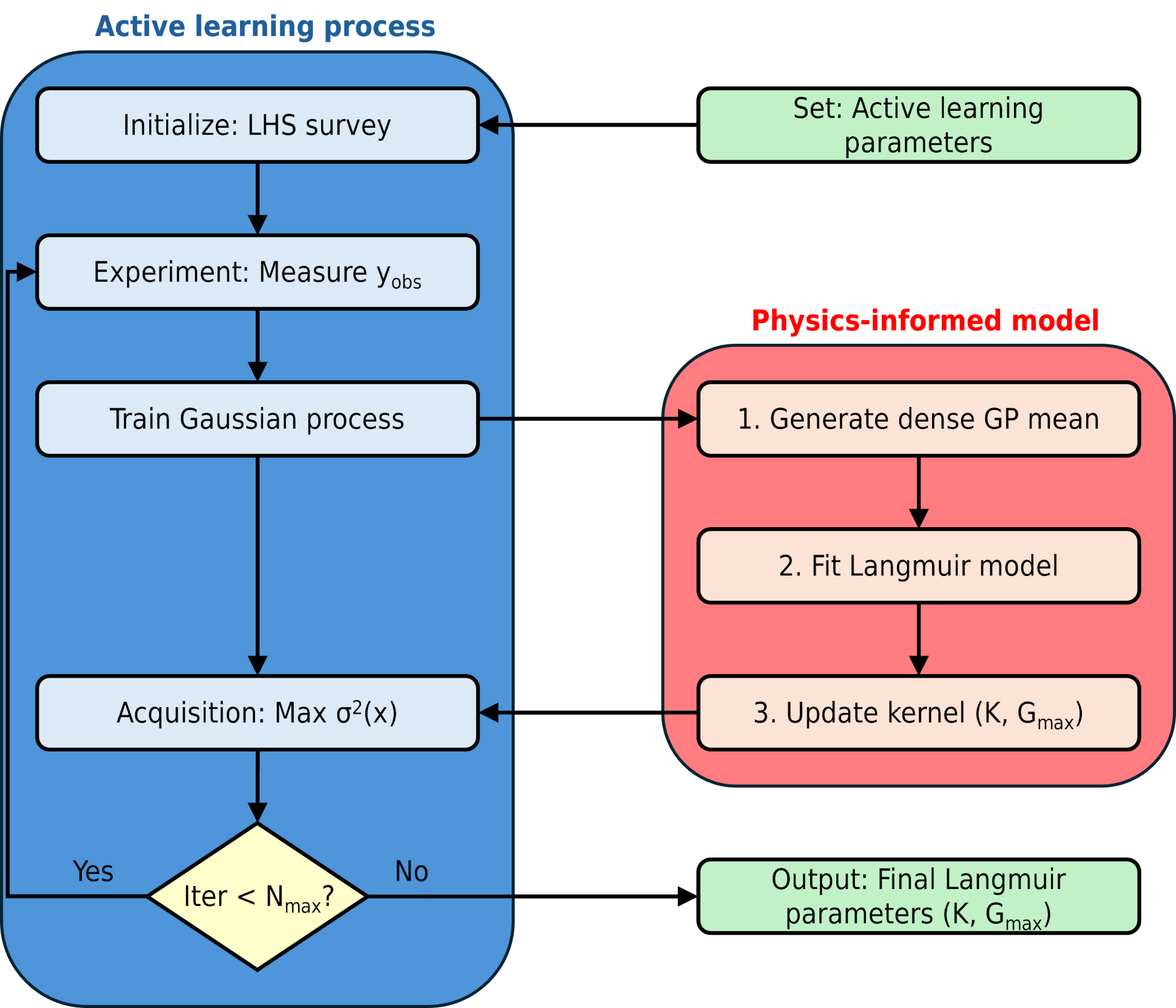}
\caption{\label{fig:flowchart}Flowchart of the Physics-Informed Bayesian Active Learning framework. The red highlighted section details the novel two-stage parameter estimation strategy, where physical parameters are extracted from the smoothed GP surrogate to ensure robustness against experimental noise.}
\end{figure}

The iterative exploration procedure, illustrated in Figure~\ref{fig:flowchart}, follows these steps:

\begin{enumerate}
\item \textbf{Initialize:} Latin Hypercube Sampling (LHS) generates $N_{\text{survey}}$ random points in the exploration window $[x_{\min}, x_{\max}]$ to initiate BAL parameters. LHS ensures better space-filling coverage than simple random sampling.
\item \textbf{Train GP:} Fit the GP to observed data $(x, y_{\text{obs}})$ with noise variance $\sigma_{\text{total}}^2(x)$.
\item \textbf{Predict:} Compute the posterior mean $\mu(x)$ and standard deviation $\sigma(x)$ over a dense grid.
\item \textbf{Explore:} Choose the next sampling point as $x_{\text{next}} = \arg\max_x \sigma^2(x)$, implementing a pure exploration (maximum uncertainty) acquisition function. Note that while this framework utilizes the machinery of BO, it functions as a BAL protocol; the objective is not to maximize the growth rate itself, but to minimize the uncertainty in the model parameters ($K$, $G_{\max}$) to accurately predict the saturation behavior.
\item \textbf{Sample:} Measure new data $y_{\text{next}}$ from the true Langmuir model with noise according to Equation~\ref{eq:obs_model}.
\item \textbf{Update:} Generate dense GP predictions over a high-resolution grid (1000 points per unit interval). Fit the Langmuir model to these GP mean predictions to obtain updated parameter estimates. For the physics-informed kernel, update $(K_{\text{phys}}, G_{\text{phys}})$ using these GP-fitted values. Calculate cumulative exploration cost ($\sum x_i$) and relative error in predicting $x_{\text{sat}}$ for target saturation.
\item \textbf{Repeat:} Continue until $N_{\text{iter}}$ iterations are reached. For statistical robustness, the full procedure is repeated over $N_{\text{MC}}$ Monte Carlo runs with independent noise realizations, and statistical outliers are removed using the Interquartile Range (IQR) method before computing summary statistics.
\end{enumerate}

After the specified number of iterations, the predicted dose time required to reach the targeted saturation is given by:

\begin{equation}
x_{\text{sat},t} = \frac{s_{\text{target}}}{K_t(1-s_{\text{target}})}
\label{eq:xsat_fit}
\end{equation}

The true value can be calculated from:

\begin{equation}
x_{\text{sat,true}} = \frac{s_{\text{target}}}{K_{\text{true}}(1-s_{\text{target}})}
\label{eq:xsat_true}
\end{equation}

To evaluate the accuracy of the predicted value, the relative error is calculated as:

\begin{equation}
\text{Error}_{\text{sat}} = 100 \times \frac{|x_{\text{sat},t} - x_{\text{sat,true}}|}{x_{\text{sat,true}}} \quad (\%)
\label{eq:error}
\end{equation}

Cumulative exploration cost is defined as:

\begin{equation}
C(n) = \sum_{i=1}^n x_i
\label{eq:cost}
\end{equation}
where $n$ is the iteration number. This metric quantifies the total experimental resource consumption (e.g., total exposure time), serving as a proxy for precursor usage efficiency.

\section{Simulation Results}

We evaluate model performance across four distinct simulated scenarios: (i) \textbf{fast saturation}, where the optimal pulse to reach the target saturation falls near the lower limit of the search window; (ii) \textbf{moderate saturation}, where the optimal pulse falls near the upper limit; (iii) \textbf{slow saturation}, where the optimal pulse extends beyond the upper limit, requiring extrapolation; (iv) \textbf{high noise}, where the measurements have intensified uncertainties. To model scenarios (i)--(iii), we maintain constant values for $G_{\max,\text{true}} = 1.0$, $x_{\min} = 0.005$, and $x_{\max} = 10.0$ while varying $K_{\text{true}}$ to position the saturation point within the respective regions. Across these scenarios, we introduce realistic experimental noise with $\sigma_{y,\text{abs}} = 0.05$ and $\sigma_{x,\text{abs}} = 0.005$, and initialize each run with $N_{\text{survey}} = 5$ LHS survey scans. The target saturation was set to 99\% of $G_{\max}$ for all simulations. We perform 20 iterations to demonstrate the capability of our model to achieve accurate predictions with relatively few iterations.

\subsection{Exploration Behavior}

\begin{figure*}
\includegraphics[width=6.75in]{Figures/Exploration.png}
\caption{\label{fig:exploration}Exploration behavior of the physics-informed Matérn kernel vs.\ pure Matérn kernel across all four simulated regimes over 20 iterations. Gray shaded regions represent 95\% confidence intervals. Colors indicate iteration number. (a) Fast saturation ($K_{\text{true}} = 100$). (b) Moderate saturation ($K_{\text{true}} = 10$). (c) Slow saturation ($K_{\text{true}} = 1$). (d) High noise ($K_{\text{true}} = 10$, $\sigma_{y,\text{abs}} = 0.1$, $\sigma_{x,\text{abs}} = 0.01$).}
\end{figure*}

In active learning for ALD optimization, where the learner chooses to sample is as consequential as what it ultimately predicts: the spatial distribution of acquired points reveals whether the algorithm is actively exploiting embedded physical knowledge or merely reducing GP uncertainty without regard for the underlying kinetics. A physics-informed kernel should concentrate acquisitions in the most informative regions of the adsorption isotherm (specifically, where the Langmuir curve exhibits the steepest slope and each new measurement most efficiently constrains the saturation parameters) rather than dispersing effort uniformly across the search window. Examining exploration behavior therefore serves two purposes: it validates that the embedded Langmuir structure is genuinely shaping the acquisition strategy, and it provides mechanistic intuition for the convergence gains quantified in the subsequent analysis. We compare the spatial distributions of sampled exposure times for the physics-informed and pure Matérn kernels over 20 iterations across all four simulated regimes.

\textbf{Fast saturation:} We set $K_{\text{true}} = 100$ to evaluate model performance when saturation occurs rapidly within the lower portion of the search window. Figure~\ref{fig:exploration}a compares the exploration behavior of the two kernel strategies. The physics-informed Matérn kernel demonstrates a marked preference for sampling regions where the adsorption isotherm exhibits a steeper slope, corresponding to the most informative portion of the Langmuir curve. This targeted exploration contrasts sharply with the pure Matérn kernel, which bases its sampling decisions solely on GP uncertainty without incorporating knowledge of the underlying adsorption kinetics.

\textbf{Moderate saturation:} We set $K_{\text{true}} = 10$ to position the saturation point near the upper limit of the search window. Figure~\ref{fig:exploration}b illustrates the exploration behavior of the two kernels. In this scenario, the physics-informed Matérn kernel demonstrates a more distributed exploration pattern compared to the fast saturation case. While the kernel still leverages the embedded Langmuir structure to guide sampling, exploration of the high-slope region is less intensive, as the saturation curve extends further into the search domain. This behavior reflects the adaptive nature of the physics-informed approach, which adjusts its exploration strategy based on the learned parameters ($K_{\text{phys}}$, $G_{\text{phys}}$) as they converge toward the true values.

\textbf{Slow saturation:} We set $K_{\text{true}} = 1$ to position the saturation point beyond the upper limit of the search window, requiring extrapolation from the explored region to predict the target saturation time. While this regime demonstrates the robustness of our framework for extrapolation and may prove useful for precursors with substantially low vapor pressure, it does not represent a typical application scenario. In standard ALD process development, practitioners select search windows that encompass the expected saturation range rather than targeting saturations far beyond the upper limit, as this would result in low-throughput processing. Figure~\ref{fig:exploration}c demonstrates the exploration behavior. In this challenging scenario, the physics-informed Matérn kernel exhibits an even more distributed exploration pattern than observed in the moderate saturation case. The kernel spreads its sampling across a broader range of exposure times, attempting to capture the gradual approach toward saturation that extends beyond the predefined search boundary. This behavior contrasts with the more localized exploration strategies observed when saturation falls within the search window, reflecting the model's adaptation to the underlying kinetics.

\textbf{High noise:} To evaluate the robustness of both kernel strategies under challenging experimental conditions, we conducted an additional analysis using substantially elevated noise levels. For this investigation, we employed the moderate saturation regime ($K_{\text{true}} = 10$), which provided the most favorable balance between exploration window coverage and extrapolation requirements in previous sections. The noise parameters were increased relative to the baseline simulations: $\sigma_{y,\text{abs}} = 0.1$ for measurement uncertainty and $\sigma_{x,\text{abs}} = 0.01$ for timing precision. Correspondingly, $x_{\min}$ was raised from 0.005~s to 0.010~s to ensure that timing jitter does not produce non-physical negative exposure values. These elevated noise levels represent scenarios where measurement equipment operates near its detection limits or where environmental fluctuations significantly impact data quality. As shown in Figure~\ref{fig:exploration}d, the substantially increased uncertainty manifests as greater scatter in the observed data points relative to the true Langmuir curve, reflecting the challenging measurement environment. Despite this heightened noise, the physics-informed Matérn kernel maintains its characteristic sampling strategy, continuing to prioritize regions where the adsorption isotherm exhibits steeper slopes. This behavior demonstrates the robustness of the physics-informed approach: even when individual measurements are significantly corrupted by noise, the embedded Langmuir structure guides the learner toward the most informative regions of the parameter space, where the signal-to-noise ratio for learning the underlying kinetics remains relatively favorable.

\subsection{Monte Carlo Analysis}

\begin{figure*}
\includegraphics[width=6.75in]{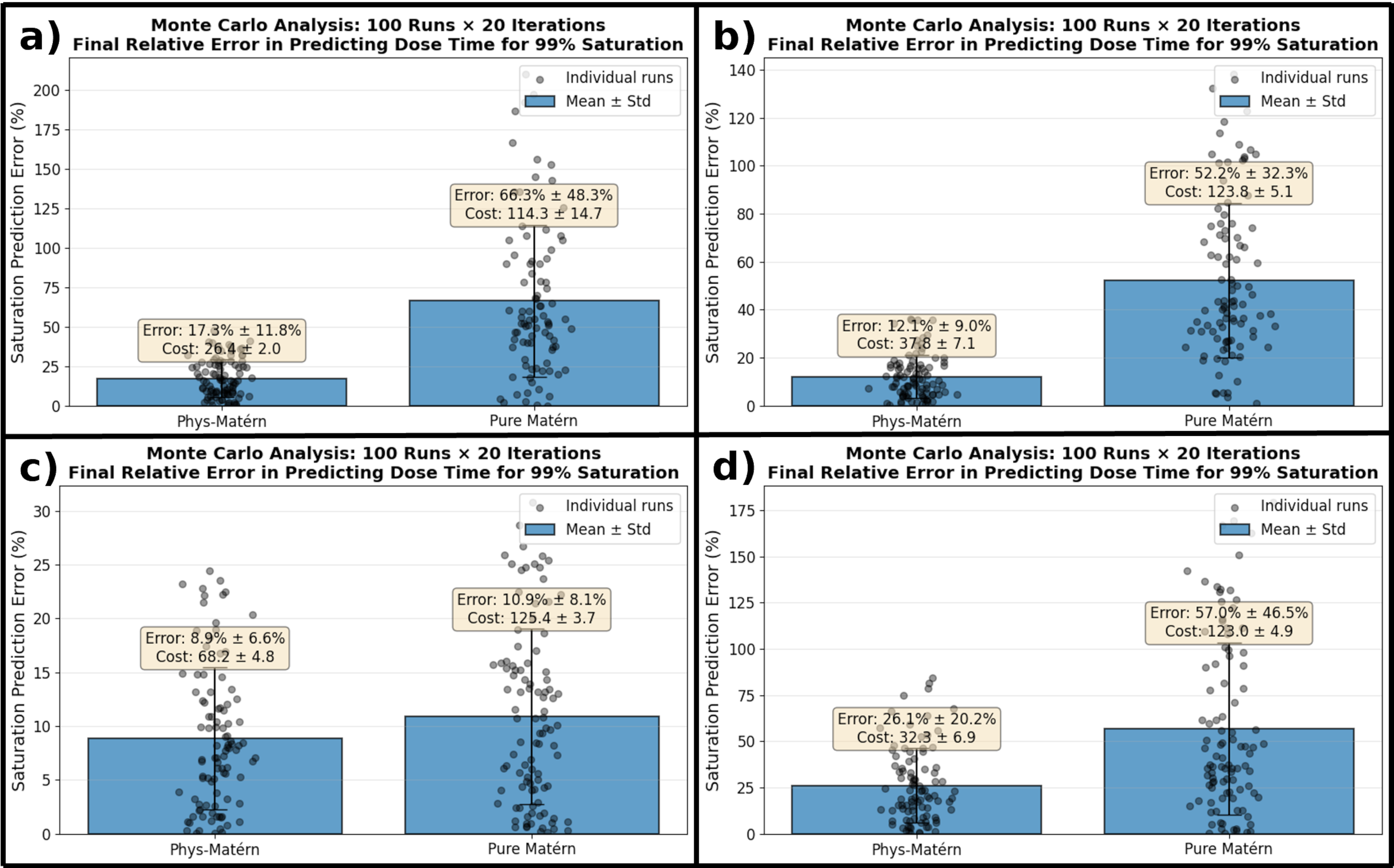}
\caption{\label{fig:mc}Monte Carlo statistical analysis (100 runs, 20 iterations) comparing prediction error and precursor usage for the physics-informed Matérn (left bar) and pure Matérn (right bar) kernels. Error bars represent mean $\pm$ standard deviation after outlier removal via IQR method. (a) Fast saturation ($K_{\text{true}} = 100$). (b) Moderate saturation ($K_{\text{true}} = 10$). (c) Slow saturation ($K_{\text{true}} = 1$). (d) High noise ($K_{\text{true}} = 10$, $\sigma_{y,\text{abs}} = 0.1$, $\sigma_{x,\text{abs}} = 0.01$).}
\end{figure*}

Single-run comparisons can be dominated by favorable or unfavorable noise realizations, making it difficult to distinguish systematic performance differences from sampling artifacts. To obtain statistically robust estimates, we performed 100 Monte Carlo simulations per regime, each with independently drawn random error realizations, and removed statistical outliers via the interquartile range (IQR) method before computing summary statistics. We report the mean prediction error for the 99\% saturation time after 20 iterations alongside the precursor usage reduction factor, comparing the physics-informed and pure Matérn kernels across all four regimes.

\textbf{Fast saturation:} Figure~\ref{fig:mc}a shows that the physics-informed kernel predicts the 99\% saturation time with a mean error of 17.3$\pm$11.8\%, compared to 66.3$\pm$48.3\% for the pure Matérn kernel, a nearly fourfold improvement in prediction accuracy. The cost evaluation reveals more than fourfold improvement in precursor usage efficiency, demonstrating substantial resource savings alongside enhanced predictive performance.

\textbf{Moderate saturation:} Figure~\ref{fig:mc}b shows that the physics-informed model predicts the 99\% saturation time with a mean error of 12.1$\pm$9.0\%, compared to 52.2$\pm$32.3\% for the pure Matérn kernel. The cost analysis demonstrates approximately threefold improvement in precursor usage efficiency, highlighting significant resource savings even when saturation occurs at longer exposure times.

\textbf{Slow saturation:} Even in this extrapolation regime, the physics-informed kernel maintains an accuracy advantage. As illustrated in Figure~\ref{fig:mc}c, the physics-informed model predicts the 99\% saturation time with a mean error of 8.9$\pm$6.6\%, compared to 10.9$\pm$8.1\% for the pure Matérn kernel, representing approximately 20\% improvement. The embedded physical structure provides meaningful guidance even when extrapolating beyond the explored region. The cost analysis reveals approximately twofold improvement in precursor usage efficiency, demonstrating dual advantages in accuracy and resource consumption.

\textbf{High noise:} Even under degraded measurement conditions, the physics-informed kernel maintains a significant performance advantage. As shown in Figure~\ref{fig:mc}d, the physics-informed model predicts the 99\% saturation time with a mean error of 26.1$\pm$20.2\%, compared to 57.0$\pm$46.5\% for the pure Matérn kernel. While both models exhibit larger prediction errors than under nominal noise conditions, the physics-informed approach achieves approximately 50\% improvement in accuracy. The cost analysis continues to demonstrate approximately fourfold improvement in precursor usage efficiency, confirming that these performance gains are achieved with more efficient resource utilization.

\subsection{Convergence Analysis}

\begin{figure*}
\includegraphics[width=6.75in]{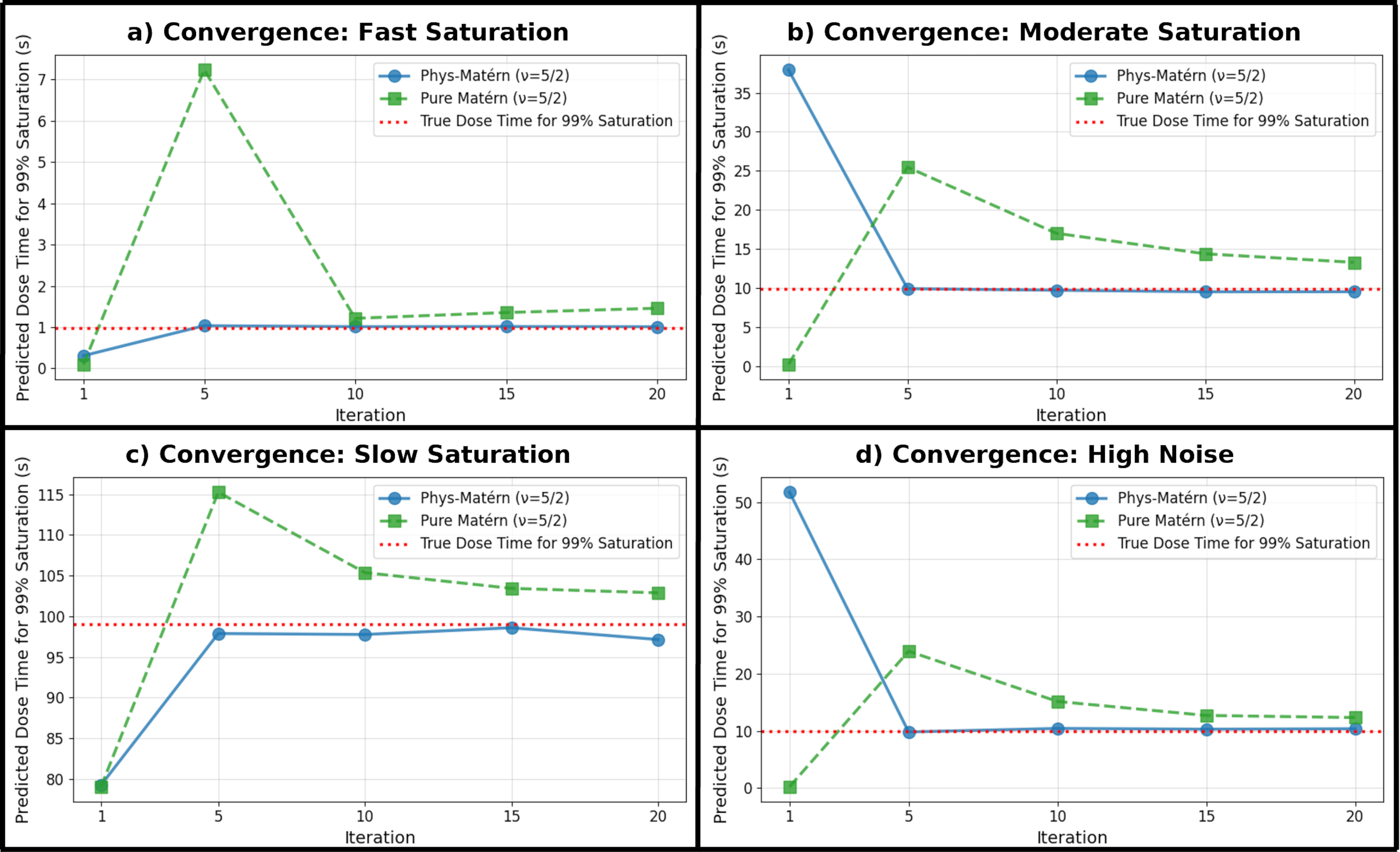}
\caption{\label{fig:convergence}Convergence of the predicted saturation time across four simulated regimes, averaged over 100 Monte Carlo runs. The red dotted line represents the true saturation time. (a) Fast saturation ($K_{\text{true}} = 100$). (b) Moderate saturation ($K_{\text{true}} = 10$). (c) Slow saturation ($K_{\text{true}} = 1$). (d) High noise ($K_{\text{true}} = 10$, $\sigma_{y,\text{abs}} = 0.1$, $\sigma_{x,\text{abs}} = 0.01$).}
\end{figure*}

In active learning for ALD optimization, each iteration corresponds to a physical deposition experiment, making rapid convergence a practical necessity: the fewer iterations required to reliably identify the saturation time, the less precursor material and process time are consumed. A method that converges slowly, or not at all within the available budget, offers limited value regardless of its asymptotic accuracy. To assess the learning efficiency of the proposed framework, we therefore examine how quickly the predicted saturation time converges to the true value across all four regimes, comparing the physics-informed and pure Matérn kernels over 20 iterations.

\textbf{Fast saturation:} Figure~\ref{fig:convergence}a shows that the physics-informed model converges to the true saturation value within five iterations, with predictions remaining stable throughout the 20-iteration budget. The pure Matérn kernel, by contrast, shows no meaningful improvement during these early steps. This confirms that embedding physical structure into the kernel not only improves prediction accuracy but also substantially accelerates convergence.

\textbf{Moderate saturation:} Similar rapid convergence is observed when the saturation point lies near the upper limit of the search window. As shown in Figure~\ref{fig:convergence}b, the physics-informed model stabilizes near the true value by iteration five, while the pure Matérn kernel requires significantly more iterations to reach a reliable estimate. This confirms that the performance advantage is robust to the position of the saturation point within the search domain.

\textbf{Slow saturation:} This regime demonstrates the framework's most notable extrapolation capability. As shown in Figure~\ref{fig:convergence}c, even though the true saturation lies outside the search space, the physics-informed model accurately predicts the saturation time within a few iterations. The pure Matérn kernel, lacking a structured prior, struggles to extrapolate reliably in the early iterations. This underscores a key advantage of the physics-informed kernel: by encoding the asymptotic behavior of the saturation curve, it enables confident extrapolation beyond the sampled region.

\textbf{High noise:} Figure~\ref{fig:convergence}d illustrates that the two-stage parameter estimation strategy is particularly effective under elevated noise conditions. Even with significant data scatter, the physics-informed model accurately predicts the saturation time within five iterations by leveraging the GP's smoothing capabilities, guided by the physical constraints of the Langmuir isotherm.

Table~\ref{tab:performance_summary} summarizes the performance metrics across all four regimes, highlighting the consistent efficiency gains and accuracy improvements offered by the physics-informed approach.

\begin{table}
\caption{\label{tab:performance_summary}Summary of performance over 20 iterations}
\begin{ruledtabular}
\begin{tabular}{lccc}
 & \multicolumn{2}{c}{\textbf{Prediction Error (\%)}} & \textbf{Precursor Usage} \\
\textbf{Regime} & \textbf{Phys-Matérn} & \textbf{Pure Matérn} & \textbf{Reduction Factor} \\
Fast & 17.3 $\pm$ 11.8 & 66.3 $\pm$ 48.3 & $\sim$4$\times$ \\
Moderate & 12.1 $\pm$ 9.0 & 52.2 $\pm$ 32.3 & $\sim$3$\times$ \\
Slow & 8.9 $\pm$ 6.6 & 10.9 $\pm$ 8.1 & $\sim$2$\times$ \\
High noise & 26.1 $\pm$ 20.2 & 57.0 $\pm$ 46.5 & $\sim$4$\times$ \\
\end{tabular}
\end{ruledtabular}
\end{table}

\section{Experimental Validation}

\subsection{Experimental Setup}

To validate our framework in a real experimental setting, we applied our physics-informed approach to predict the saturation time for Tetrakisdimethylamido Titanium (TDMAT) in TDMAT/\ce{O3} ALD for \ce{TiO2} deposition on silicon wafers using a commercial system (ALD-150LX, Kurt J. Lesker Company). The experimental parameters were as follows: substrate temperature of 150~$^\circ$C (selected to avoid precursor decomposition while ensuring sufficient reactivity), chamber pressure of 1~Torr, TDMAT canister temperature of 50~$^\circ$C, and precursor line heating to 100~$^\circ$C to prevent condensation.

The pulse/purge sequence was defined as $(x : 30 : 1 : 15) \times 4 + (x : 30 : 1 : 600)$ where $x$ represents the tuned TDMAT pulse time, followed by a 30-second Ar purge, a 1-second \ce{O3} pulse (15\% nominal concentration), and a final Ar purge. This sequence consists of four stabilization cycles with standard 15-second purges to reach equilibrium, followed by a fifth measurement cycle with an extended 600-second purge. Growth per cycle (GPC) was determined by measuring the thickness difference for a single pulse inserted after four consecutive pulses, specifically comparing measurements taken 15 seconds after the final purge and immediately before the fifth precursor pulse. This five-pulse protocol combined with the 600-second purge is critical for handling non-ideal desorption effects. The extended purge minimizes desorption between measurements, while the four-pulse conditioning ensures that any residual desorption appears as a reproducible offset rather than a time-varying artifact that would distort the Langmuir kinetics significantly. Thickness measurements were performed using in situ ellipsometry (FS-1EX, FilmSense) at a 70$^\circ$ incident angle with a predefined Si/\ce{SiO2}(native oxide)/\ce{TiO2} optical model in the provided software.

\subsection{Experimental Results and Discussion}

Using a modified experimental version of the code to get growth input directly from user, we conducted five survey scans followed by 10 BAL iterations. For the experimental runs, the search window was defined as 0.200--5.000 seconds. This lower bound (0.200~s) differs from the simulation lower bound (0.005~s) because the effect of the desorption term becomes significantly greater relative to the growth at very short pulse times. In this region, deviations from ideal Langmuir behavior increase, which can negatively affect the acquisition function and isotherm fitting and cause a significant deviation between GP fit and ideal Langmuir fit at the extreme lower end. By restricting the window to 0.200~s, we mitigate the influence of these desorption artifacts on the learning and prediction process.

The experimental validation was conducted exclusively with the physics-informed model and does not include a parallel comparison against the pure Matérn (non-physics-informed) approach. This decision was deliberate: the extensive simulation study presented in Section~III consistently demonstrated that the physics-informed kernel outperforms the pure Matérn kernel across all four regimes, achieving up to fourfold lower prediction error and fourfold reduction in precursor usage. Repeating the full experimental campaign with the non-physics-informed model would require a comparable number of ALD depositions at additional cost and time, without offering new scientific insight beyond what the simulations already establish. We therefore proceeded directly with the superior method for experimental validation, reserving the comparative analysis for the in silico study where it can be conducted rigorously over many Monte Carlo runs.

\begin{figure}
\includegraphics[width=3.375in]{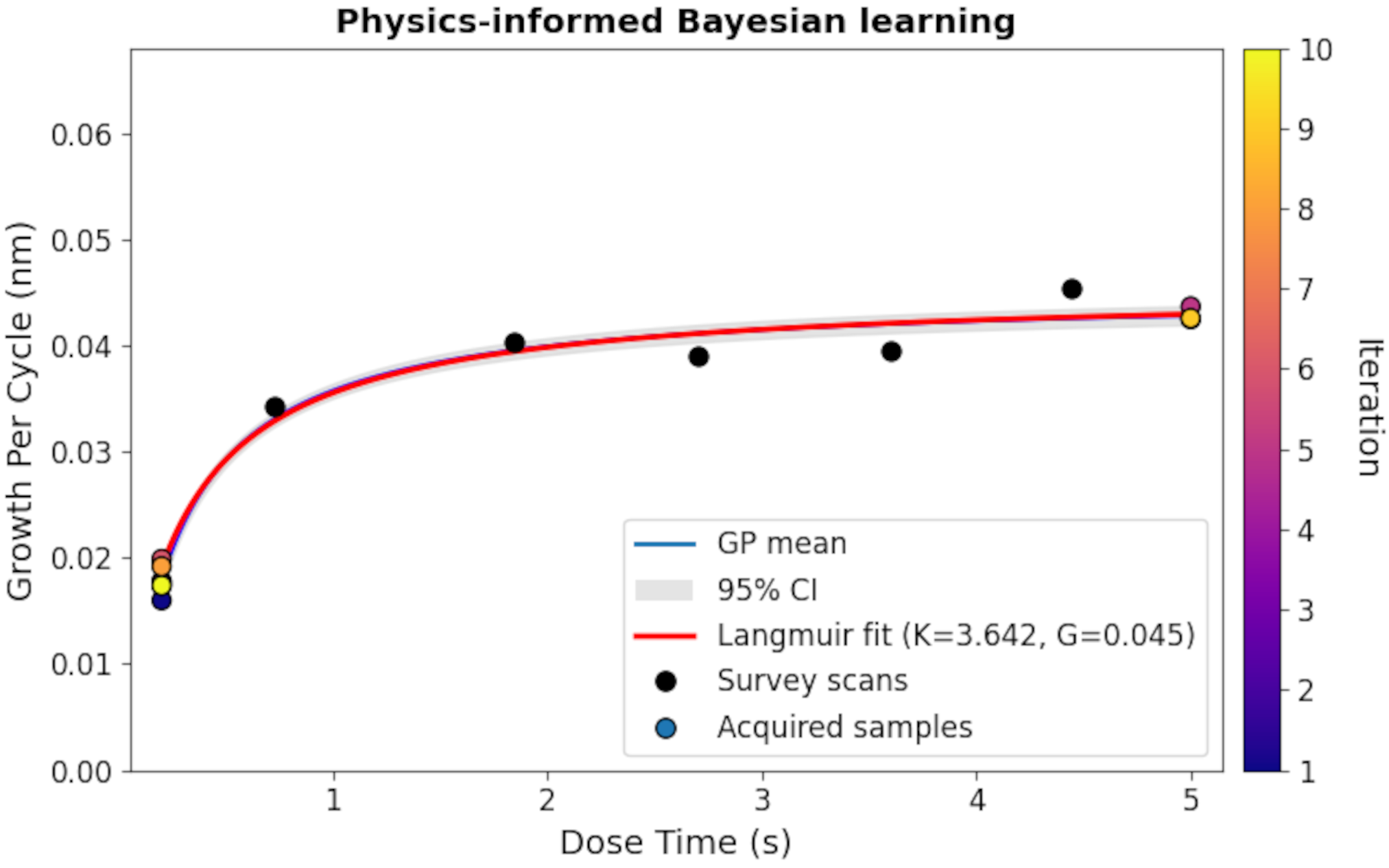}
\caption{\label{fig:exp_sampling}Distribution of sampled points across iterations during the experimental validation of TDMAT/\ce{O3} ALD. The Bayesian framework concentrates sampling in the high-saturation regime and steep-slope region while utilizing survey points for mid-range coverage. Colors indicate iteration number.}
\end{figure}

\begin{figure}
\includegraphics[width=3.375in]{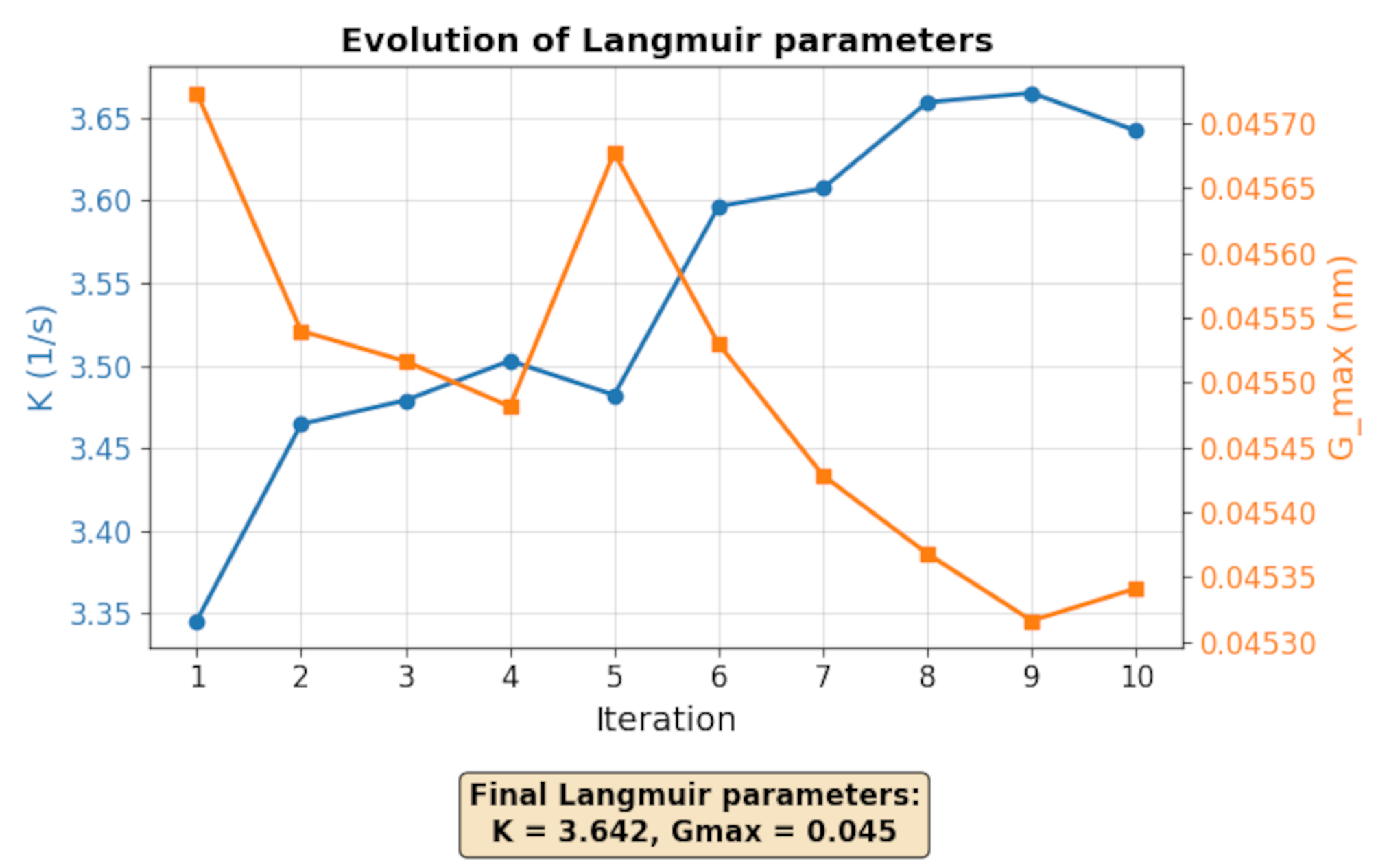}
\caption{\label{fig:exp_results}Experimental results showing the evolution of Langmuir parameters $K$ and $G_{\max}$ across active learning iterations. Both parameters demonstrate stable convergence, validating the two-stage parameter estimation strategy.}
\end{figure}

Figure~\ref{fig:exp_sampling} shows the distribution of sampled points across iterations. Notably, after the initial survey scans, the Bayesian framework predominantly selected experimental points in two regions: the highly saturated regime (highest pulse time) and the steepest portion of the isotherm where uncertainty is greatest, while relying on survey scans for the mid-range. This sampling strategy is consistent with the behavior observed in simulations and reflects the physics-informed kernel's bias toward informative regions. Figure~\ref{fig:exp_results} presents the experimental results and the evolution of $K$ and $G_{\max}$ values across iterations, demonstrating stable convergence of both physical parameters.

\begin{figure*}
\includegraphics[width=6.75in]{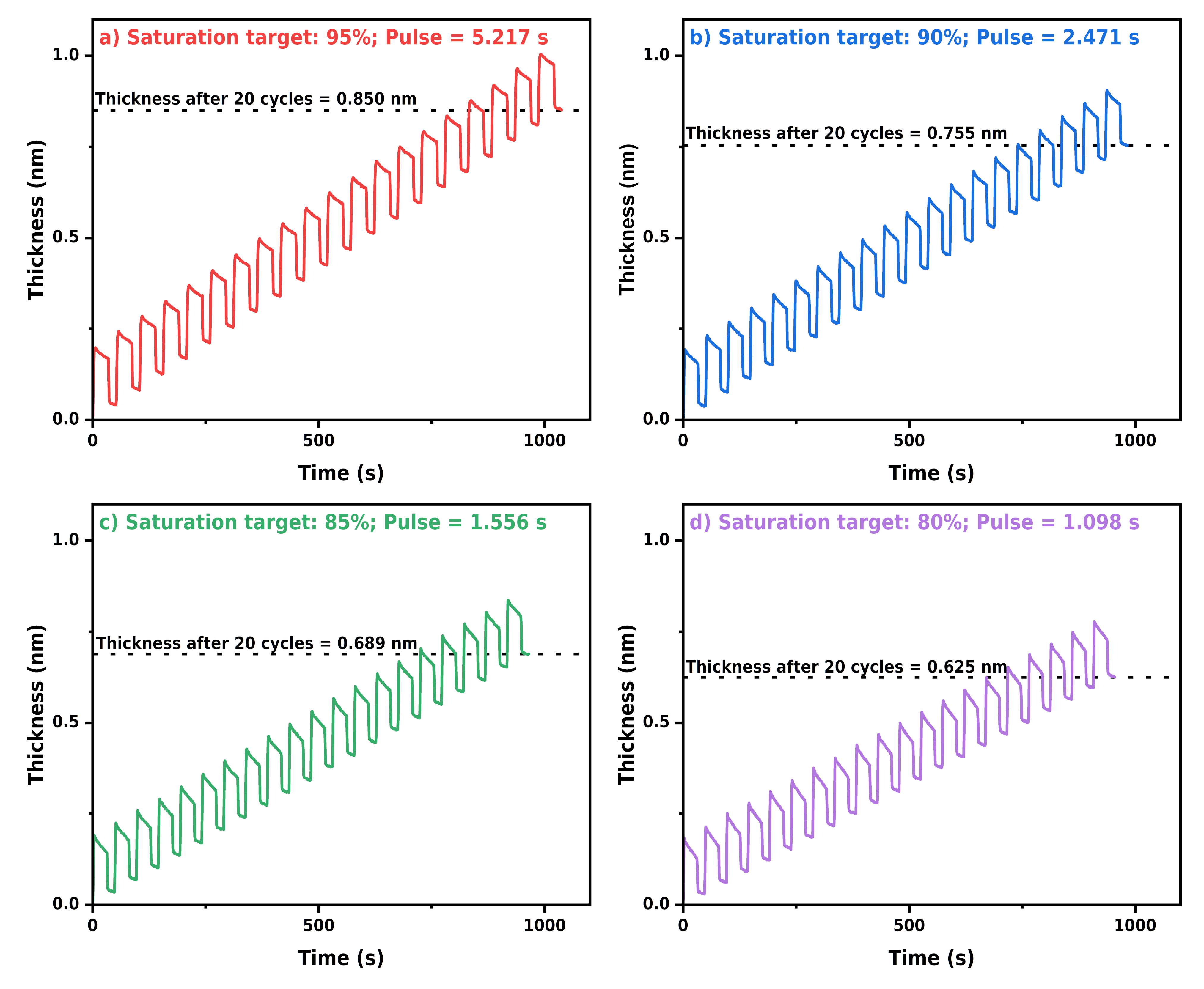}
\caption{\label{fig:exp_accuracy}Accuracy assessment of predicted saturation times. The plots show the growth over 20 cycles for four different pulse times: (a) 95\% saturation (5.217~s, GPC = 0.0425~nm), (b) 90\% saturation (2.471~s, GPC = 0.0378~nm), (c) 85\% saturation (1.556~s, GPC = 0.0345~nm), and (d) 80\% saturation (1.098~s, GPC = 0.0313~nm). The maximum GPC ($G_{\max}$) is 0.045~nm. While the model accurately predicts high saturation targets (95\%), deviations increase at lower targets due to non-ideal adsorption and desorption effects.}
\end{figure*}

To assess the accuracy of the predicted saturation times, we evaluated the experimental growth at targeted saturation levels of 0.95, 0.90, 0.85, and 0.80 by using the final parameters obtained from BAL shown in Figure~\ref{fig:exp_results} and using Equation~\ref{eq:xsat_fit} to calculate respective pulse times. The results are summarized in Table~\ref{tab:exp_results_error} and visualized in Figure~\ref{fig:exp_accuracy}. The maximum growth per cycle ($G_{\max}$) was determined to be 0.045~nm, which matches reported growth for TDMAT/\ce{O3} ALD in the literature.\cite{29} The model accurately predicted the saturation time for the 0.95 saturation target (5.217~s), yielding a GPC of 0.0425~nm (0.6\% error). However, as the targeted saturation decreased to lower values, the deviation between model predictions and experimental results increased. For the 0.90 target (2.471~s), the GPC was 0.0378~nm (6.8\% error); for the 0.85 target (1.556~s), the GPC was 0.0345~nm (9.8\% error); and for the 0.80 target (1.098~s), the GPC was 0.0313~nm (13.0\% error). 

\begin{table*}
\caption{\label{tab:exp_results_error}Comparison of targeted vs. measured saturation levels and prediction error}
\begin{ruledtabular}
\begin{tabular}{ccccc}
\textbf{Targeted Saturation (\%)} & \textbf{Pulse Time (s)} & \textbf{GPC (nm)} & \textbf{Measured Saturation (\%)} & \textbf{Error (\%)} \\
95 & 5.217 & 0.0425 & 94.4 & 0.6 \\
90 & 2.471 & 0.0378 & 84.0 & 6.7 \\
85 & 1.556 & 0.0345 & 76.7 & 9.8 \\
80 & 1.098 & 0.0313 & 69.6 & 13.0 \\
\end{tabular}
\end{ruledtabular}
\end{table*}

We attribute this discrepancy to the desorption term $D(x)$ in Equation~\ref{eq:langmuir_desorption}. While the extended 600-second purge minimizes this effect, it cannot be entirely eliminated, and its relative impact is more pronounced at lower surface coverages where the driving force for adsorption is lower. At high saturation levels (95\%), the surface is nearly fully occupied, and non-ideal effects represent a small fraction of the total coverage. At lower saturation levels, however, fewer sites are occupied, and when a long pulse follows by a short pulse without extended purge, the same amount of continuous desorption results in a larger relative loss when equilibrium within the cycles is not achieved. This behavior is consistent with the non-ideal kinetics expected in real ALD systems and highlights the importance of considering alternative kinetic models when targeting intermediate and low coverage regimes.

\section{Conclusion}
We have demonstrated a physics-informed BAL framework that significantly accelerates the determination of saturation pulse times in ALD. By embedding the Langmuir isotherm, the most widely accepted and fundamental model for ALD saturation behavior directly into the covariance kernel and utilizing a GP-smoothed parameter estimation technique, the model achieves rapid convergence even in the presence of significant measurement noise and timing jitter. The primary goal of this work is to establish a robust active learning framework rather than to develop comprehensive kinetic models. Our simulations reveal three key advantages over standard data-driven approaches: \textbf{Efficiency:} The physics-informed model consistently identifies target saturation times within five iterations, while reducing precursor consumption by up to 75\%. \textbf{Extrapolation:} Unlike standard GPs, which struggle at the boundaries of the search space, the Langmuir-transformed kernel enables accurate extrapolation to saturation times beyond the explored experimental window. \textbf{Robustness:} The two-stage fitting strategy effectively filters experimental noise, maintaining high predictive accuracy even when signal degradation occurs.

Experimental validation using TDMAT/\ce{O3} deposition confirms the practical utility of the framework. The model successfully identified saturation times for high-coverage targets ($\geq$95\%) with less than 1\% error, validating its effectiveness for practical optimization. Our experimental observations reveal that while the Langmuir model captures the general saturation trend, it may not fully represent the true kinetics, particularly in low-saturation regimes where deviations become apparent. Despite these deviations, the framework demonstrates reliable performance for practical optimization tasks.

We recognize that a more sophisticated kinetic model (potentially incorporating more than two adjustable parameters) would better describe both classical Langmuir behavior and non-ideal low-saturation regimes. However, such a study requires extensive experimental characterization and mechanistic investigation of adsorption-desorption behaviors, which falls outside the scope of this work. To maintain focus on the active learning algorithm and its practical implementation, we defer this detailed kinetic analysis to future studies. We are currently conducting in-depth experimental characterization to develop such a model specialized to be used in our framework, which will be published separately. Importantly, the provided open-source code is modular and easily adaptable, allowing researchers to implement alternative kinetic models tailored to their specific ALD systems. This methodology provides a generalizable, sample-efficient strategy for ALD process development, offering a rigorous path toward autonomous, closed-loop tuning in experimental reactor environments.

\section*{Supplementary Material}
See supplementary material for the Python implementation code of the physics-informed Bayesian active learning framework.

\section*{acknowledgments}
The authors acknowledge financial support from University of Illinois Chicago Department of Biomedical Engineering.

\section*{Author Declarations}
\subsection*{Conflict of interest}
The authors have no conflicts to disclose.

\section*{Data Availability Statement}
Python implementation code for the physics-informed Bayesian active learning framework is provided in the supplementary material. Additional data including raw experimental measurements are available upon request from the corresponding author.

\section*{References}
\bibliography{references}

\end{document}